\documentclass[sigconf]{acmart}
\usepackage{soul}
\usepackage{xcolor}
\usepackage{booktabs}
\usepackage{amsmath}
\usepackage{algorithm}
\usepackage[noend]{algpseudocode}
\usepackage{subcaption}

\renewcommand\footnotetextcopyrightpermission[1]{} 
\setcopyright{none}

\begin{document}


\title{Beyond the Trees:\\Resilient Multipath for Last-mile WISP Networks}

\author{Bilal Saleem}
\affiliation{%
  \institution{New York University}
  \city{Abu Dhabi} 
}
\email{mbs10@nyu.edu}

\author{Paul Schmitt}
\affiliation{%
  \institution{Princeton}
}
\email{pschmitt@cs.princeton.edu}

\author{Jay Chen}
\affiliation{%
  \institution{New York University}
  \city{Abu Dhabi} 
}
\email{jchen@cs.nyu.edu}

\author{Barath Raghavan}
\affiliation{%
  \institution{USC}
}
\email{barath.raghavan@usc.edu}

\begin{abstract}

Expanding the reach of the Internet is a topic of widespread interest today. Google and Facebook, among others, have begun investing substantial research efforts toward expanding Internet access at the edge. Compared to data center networks, which are relatively over-engineered, last-mile networks are highly constrained and end up being ultimately responsible for the performance issues that impact the user experience.

The most viable and cost-effective approach for providing last-mile connectivity has proved to be Wireless ISPs (WISPs), which rely on point-to-point wireless backhaul infrastructure to provide connectivity using cheap commodity wireless hardware. However, individual WISP network links are known to have poor reliability and the networks as a whole are highly cost constrained.

Motivated by these observations, we propose Wireless ISPs with Redundancy (WISPR), which leverages the cost-performance tradeoff inherent in commodity wireless hardware to move toward a greater number of inexpensive links in WISP networks thereby lowering costs. To take advantage of this new path diversity, we introduce a new, general protocol that provides increased performance, reliability, or a combination of the two.

\end{abstract}

\settopmatter{printacmref=false, printfolios=true}
\maketitle

\section{Introduction}

Despite many advances in networking research, around half of the world's population remains disconnected from the Internet~\cite{worldinternetstats}. Even in a wealthy nation such as the United States, millions of people living in rural areas remain disconnected due to high connectivity costs. Google~\cite{loon}, Facebook~\cite{fbdrone}, and others~\cite{elon} have recently begun tackling universal Internet access by developing new aerial access technologies, yet many important problems in this new domain remain to be solved. To connect to users that are spread across large and hard-to-reach geographic areas, point-to-point wireless remains the most cost-effective last-mile technology. Operators of Wireless ISPs (WISPs), which connect millions of rural users around the world, rely on cheap commodity networking hardware and fragile tree-like network topologies to keep costs down. Although this hardware has improved dramatically from a decade ago---when network operators resorted to building their own makeshift gear using indoor APs---performance, reliability, and cost (per customer) in WISP networks are still much worse than in other settings (e.g., wireline ISPs, datacenters).

In this paper, we propose a new architecture for WISP networks called WISPR that improves WISP performance and reliability with little to no cost overhead in the worst case and significant cost savings in the best case. WISPR is markedly different from deployed architectures for traditional networks (datacenters, ISPs, enterprise networks) due to two key challenges unique to WISP networks. First, unlike enterprise WiFi deployments, WISPs introduce multi-hop challenges at WAN latencies and experience high variability in loss rate and availability, which we quantify in Section~\ref{sec:measure}. Consequently, capacity planning is difficult because network capacity is affected by weather and other external factors. Due to the unpredictability of links, a performant WISP protocol must be able to make use of partial availability.
Second, WISP networks must be ultra-low cost. Unlike conventional networks, WISP networks are rural, remote, and constrained in the number of users; introducing more, expensive hardware is often a poor business proposition for the WISP operator.

Wireless networks in general benefit from being adaptive in order to cope with PHY and MAC variability; in WISP networks, unpredictability is the norm, and compounds across many wireless hops.
Our architecture has two key parts to cope with challenges unique to the WISP context.  The first is a change to the network topology itself to provide more redundancy and increase path diversity. However, links must be added carefully so as to not increase the cost of the network. We show in Section~\ref{sec:price} that there is a stable quadratic tradeoff between cost and performance that can be leveraged to extend a WISP network with additional links and paths. Adding multipathing and redundancy over these paths provides even better performance than higher cost (but more reliable) links. Our analysis on a real WISP topology indicates that the WISPR approach can decrease cost by over 30\% while providing the same network capacity, or increase capacity by a $5 \times$ factor while keeping cost the same.

In Section~\ref{sec:design} we discuss how existing protocols, including multipath protocols such as MPTCP, are only able to meet a few---not all---of the requirements of the WISP setting: a flexible tradeoff of performance and reliability, zero added latency, incremental deployability, low cost, zero needed modifications to edge/user devices, asymmetric parameterization, and rapid adaptation.
The second component of WISPR is thus a new suite of adaptive protocols that operate on the WISP backbone to take advantage of newly-introduced links and paths.
WISPR is different from traditional wireless protocols that aim to improve the performance and/or behavior of individual wireless links; indeed, WISPR is necessarily agnostic to the low-level wireless characteristics of links as low-cost commodity hardware is often difficult for WISP operators to modify.

We begin in Section~\ref{sec:measure} by examining the topological and traffic characteristics of an operational WISP network that members of our team built in Northern California. We analyze 6 months of measurement data from this network and highlight problems typical of WISP networks: flaky and low-capacity network links, fragile network topologies, and significant environmental interference.
Building on our measurement study, in Section~\ref{sec:price} we examine the price and performance of commodity hardware and we make the case that there is are clear benefits to replacing single, expensive, high-performance links with many inexpensive, medium-performance links.
In Section~\ref{sec:design} we discuss why existing approaches are inadequate to take advantage of our modified WISP topologies and detail the design of WISPR's backhaul-based multipath protocols. We evaluate WISPR in Section~\ref{sec:eval} and demonstrate how, in the context of WISP networks, WISPR outperforms off-the-shelf multipath protocols such as MPTCP in terms of goodput at higher link loss rates (i.e., 15\%, 20\%) as well as consistently providing lower average delay in the presence of packet loss.

\section{Measurements and Cost Simulations}
\label{sec:measure}
Wireless ISP (WISP) networks have been little studied in the academic literature aside from a handful of studies of testbeds~\cite{patra2007wildnet} and theoretical evaluations of performance optimization~\cite{jaldimac}.  Nevertheless, they provide a significant means of connectivity in rural regions around the world, with millions of users worldwide connected to the Internet via WISPs~\cite{WISPA}. Moreover, WISPs provide connectivity in hard-to-reach places that have traditionally been difficult to serve via any other terrestrial means. Despite recent efforts by Google and Facebook, among others, to provide aerial connectivity, these same companies are simultaneously partnering with and funding WISPs to deliver connectivity.\footnote{The WISP work we describe in this section was funded by Google.}

In 2014 members of our group began the process of deploying a production WISP network to serve a rural region of Northern California (Mendocino County) that had at the time no options for broadband Internet service other than limited satellite coverage.\footnote{Commercial satellite offerings are often not even considered broadband service by the definitions of regulators because the latency of such satellite service is too high.}  In addition, we studied the WISP network ecosystem and sought to understand the challenges that WISP network operators face as they build, expand, and maintain their networks~\cite{celerate2015}.  Through our own experience building a production WISP network, one that today is the primary Internet provider for the majority of households in the region, we found that standard protocols do not meet the needs of WISP operators.  This applies both to standard protocols available in commodity devices and to research approaches in the literature.  Specifically, WISP networks typically run on low-end commodity wireless hardware that, due to the hardware and the conditions, can suffer frequent outages: hardware glitches, wireless interference, power fluctuation, firmware bugs, etc. Building a reliable network infrastructure on top of these unreliable components is a challenge no longer found in production quality networked systems (e.g. datacenter networks or ISPs) that have service contracts with customers.

\subsection{Network characteristics}

\begin{figure}[t!]
	\centering
  	\includegraphics[width=0.6\columnwidth]{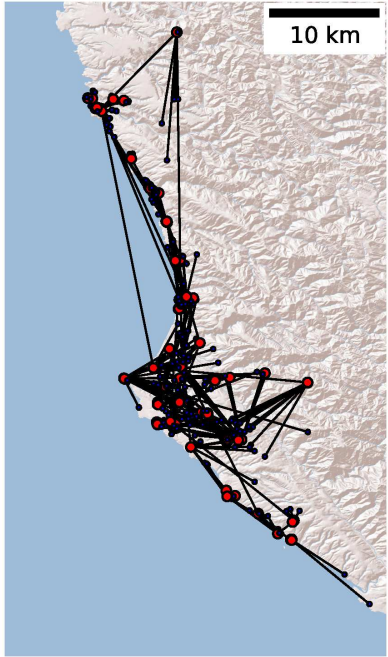}
  	\caption{WISP network topology. Red vertices indicate backhaul nodes, blue vertices indicate customer-premises equipment.}
  	\label{fig:topo}
\end{figure}

A map of the network we built is shown in Figure~\ref{fig:topo}. As is common with WISP networks, the network topology is mostly a tree. Larger red vertices in the figure indicate network backhaul devices; smaller blue vertices mark nodes located at end-user locations. All wireless devices operate on the 5GHz WiFi frequency band. 
The network uses commodity hardware from vendors like Ubiquiti~\cite{ubiquiti} and Mikrotik~\cite{mikrotik}.  The devices are almost all directional, point-to-point or point-to-multipoint wireless hardware using unlicensed or lightly-licensed spectrum.  As line of sight is required for this type of wireless connectivity, we take advantage of natural topographical advantages so as to avoid building many towers. Some sites are powered partially or exclusively using solar or wind power, and many sites have battery backup, as the power grid in the region is not reliable. The region also tends to have extreme weather during the winter months, with high winds and heavy rainfall, especially on the slopes of the densely-forested coastal mountains across which our sites are spread.

\subsubsection{Link distances}

\begin{figure}[t!]
	\centering
  	\includegraphics[trim=0 0 0 0, clip=true,width=.75\columnwidth]{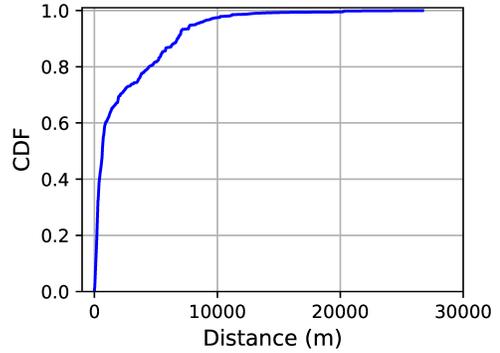}
  	\caption{Network link distance CDF.}
  	\label{fig:distances}
\end{figure}

The network includes point-to-point wireless links that span a broad range of distances. Figure~\ref{fig:distances} displays a cumulative distribution function (CDF) of link distances in meters. The median link distance is 640.9m, the mean is 2,210.9m, and the longest wireless link in the network is 26,698.4m. Because of the wide range of distances, we anticipated that link stability and performance vary (e.g., links that span longer distance are likely more susceptible to weather-related factors).

\subsubsection{Link capacities}

\begin{figure}[t!]
	\centering
  	\includegraphics[trim=0 0 0 0, clip=true,width=0.75\columnwidth]{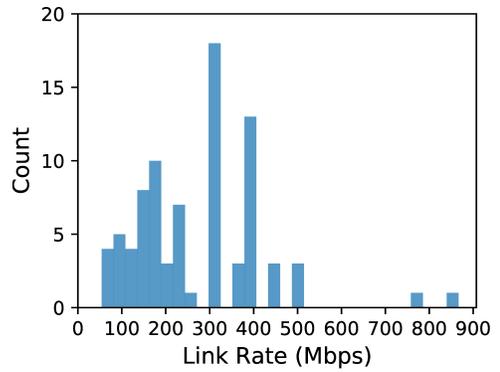}
  	\caption{Distributions of network link capacities.}
  	\label{fig:txdist}
\end{figure}

The WISP includes heterogeneous wireless hardware supporting a wide range of link capacities. Figure~\ref{fig:txdist} shows a histogram of the wireless link capacities as reported through SNMP responses. We observe that generally network backhaul links operate at less than 400 Mbps. We tracked the advertised link capacities for each node over time, as signal degradation can lead the hardware to renegotiate at lower speeds, and found that the rates remained within roughly 10\% of the rates in the figure during the observation period.

\subsubsection{Spectrum availability}

When leveraging WiFi-based equipment to construct a network, particularly as we intend to add additional and redundant links to our topology, we must consider wireless spectrum availability at any given site. Importantly, WISP network backhaul links typically rely on highly-directional antennas to form long-distance links. Raman found that co-located wireless devices using directional antennas can avoid interference by employing a 30\textdegree~\cite{Raman:2005:DEN:1080829.1080847, Nedevschi:2008:AHP:1409944.1409974} angular separation. Additionally, the 5GHz wireless band contains many non-overlapping channels to use (e.g. 24 20MHz-wide channels and 11 40MHz-wide channels). Given these factors, we anticipate that link multiplicity is achievable in the majority of cases, as we explore later in this section.

\subsection{Network measurement campaign}

To gain insight into the performance and variability of WISP network links, we collected SNMP network statistics from all core wireless nodes (i.e., excluding user devices) every 5 minutes from September 2016 through March 2017. We also collected a smaller set of data from a second WISP network that shows similar characteristics.

\subsubsection{Traffic load}

We examine the traffic load on each of the network backhaul nodes by reporting the total packets delivered. We group backhaul nodes based on the number of their descendants in the network tree topology; the mean for each group is shown in Table~\ref{tab:traffic}. As expected, nodes nearer the root (the internet gateway) generally carry more traffic than nodes with fewer descendants.  However, we find that these links do not have substantially more capacity than links further down the tree.

\begin{table}[h]
\centering
\begin{tabular}{|c|c|}
\hline
\textbf{Number of descendants} & \multicolumn{1}{c|}{\textbf{Mean packets per day}} \\ \hline
\textless 5                    & 61,073                                            \\ \hline
5 - 10                         & 20,266                                            \\ \hline
10 - 20                        & 66,448                                            \\ \hline
20 - 30                        & 168,368                                           \\ \hline
30 - 40                          & 233,680                                           \\ \hline
40 - 50                          & 345,679                                           \\ \hline
\textgreater50                 & 472,642                                           \\ \hline
\end{tabular}
\caption{Network traffic load in mean packets per day. Nodes are grouped by the number of descendants in the tree.}
\label{tab:traffic}
\end{table}

\subsubsection{Impact of weather on signal quality}
\begin{figure}[t!]
	\centering
  	\includegraphics[trim=0 0 0 0, clip=true,width=\columnwidth]{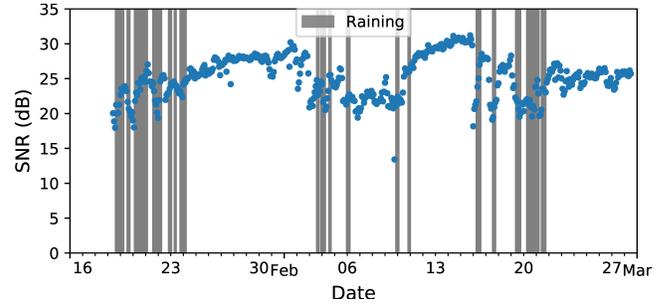}
  	\caption{Signal-to-noise ratio (SNR) for a node in the Mendocino network. Weather events result in dramatic differences in signal strength.}
  	\label{fig:snr}
\end{figure}

To illustrate the impact of rain and fog on wireless links, we plot the signal-to-noise ratio (SNR) for a single link in the network over two months in early 2017 in Figure~\ref{fig:snr}. We shade periods of time during which rain was detected at a weather station near the network in order to observe weather effects. As shown, the link experienced fluctuations as large as 12dB during rain storms. For reference, a difference of 10dB can be interpreted as a 10$ \times $ increase in signal strength. As such, weather can drastically alter the signal quality of individual wireless links. These dramatic variations make it difficult to accurately provision WISP networks and require protocols to adapt quickly to network changes.

\begin{figure}[t!]
	\centering
  	\includegraphics[trim=0 0 0 0, clip=false,width=\columnwidth]{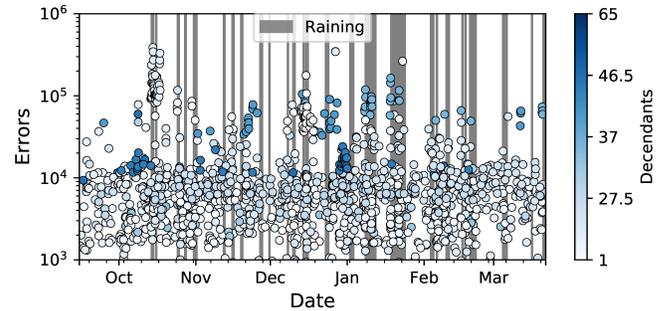}
  	\caption{Wireless node packet errors summed into 2-hour bins over the observation period. Nodes are shaded based on the number of descendant nodes in the network tree topology.}
  	\label{fig:errors_descendants}
\end{figure}
Figure~\ref{fig:errors_descendants} displays a time series from late 2016 to early 2017 with the wireless interface packet error counts for each backhaul node aggregated in two-hour increments. For clarity, we restrict the plot to outlier error counts for each node, which we define as totals greater than 3.5 standard deviations from the node's mean error count. Nodes are shaded to indicate the number of descendants in the network topology, with the root of the tree being the internet gateway. Thus, darker nodes can be interpreted as ``more important'', as errors on such links impact internet traffic for all downstream clients. We observe high variability in packet errors both between different backhaul nodes as well as between different timestamps of the same node. Such performance is different from datacenter networks and traditional ISP networks, but typical of point-to-point wireless links operating in unlicensed, WiFi-based frequencies~\cite{patra2007wildnet, Chebrolu:2006:LLP:1161089.1161099}. Furthermore, unlike edge WiFi networks, packet errors are exacerbated by the long links and multi-hop topology found in WISP networks.

A challenge unique to networks based on wireless technologies is the susceptibility of links to environmental factors such as rain or fog. Gray windows in the figure represent hours during which a weather station nearest to the network\footnote{\url{https://www.wunderground.com/personal-weather-station/dashboard?ID=KCAMANCH1}} recorded rainfall totals greater than zero. As shown, rain appears to have a variable, oftentimes substantial, impact on the error rates for some backhaul nodes. For instance, two rain storms in mid-January result in multiple nodes that experienced sharp increases in packet errors during the rain events. The effect of rain on packet errors is not universal; we posit that this could be due to differing link distance, RSSI, and location as nodes are located up to tens of kilometers from the weather station and may have experienced different weather (e.g., no rain) during the observation period.

\subsection{Price / performance analysis}
\label{sec:price}

\begin{figure}[t]
	\centering
  \includegraphics[width=0.8\columnwidth]{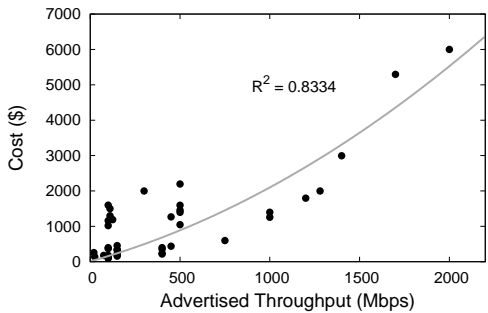}
  \vskip 1em
	\caption{Throughput vs. Cost per WISP link.}
  \vskip -1em
	\label{fig:throughput_all}
\end{figure}

Unlike traditional ISPs, WISPs often have little or no wireline infrastructure. Instead, WISPs use directional, point-to-point wireless equipment operating in unlicensed spectrum, which can increasingly offer network capacities rivaling wireline infrastructure.
Figure~\ref{fig:throughput_all} is a plot of the price and advertised throughput for a single link for various commodity wireless hardware using prices, as of July 2016, from a popular reseller\footnote{\url{http://www.streakwave.com/}} of such equipment; price represents equipment for two sides of a wireless link. We find that prices increase in a roughly quadratic manner. The figure and trend line are for all wireless equipment we found from the reseller's website; however, if we limit the equipment list to two popular vendors, Ubiquiti and MikroTik, the trend line has an $R^2$ value of 0.870, indicating a tighter fit to the data. Historical price data that we examined from the past five years appear to maintain this general relationship despite shifts in technology and speeds.

\begin{figure}[t]
	\centering
	\includegraphics[width=0.8\columnwidth]{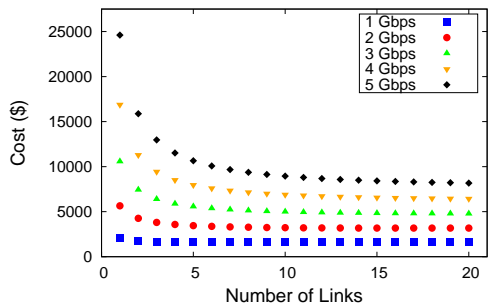}
  \vskip 1em
	\caption{Cost to deploy a link with set capacities using a varying number of underlying parallel links. The use of multiplicity reduces the capital cost.}
  \vskip -1em
	\label{fig:multiplicity}
\end{figure}

These findings suggest that network operators could potentially decrease capex if the capability existed to leverage multiple, parallel links in a WISP setting. We investigate this intuition by using the formula for the trend line in Figure~\ref{fig:throughput_all} to extrapolate the cost for wireless links of arbitrary capacities. Figure~\ref{fig:multiplicity} shows the cost to build wireless links of capacities ranging from 1 Gbps to 5 Gbps using a increasing number of parallel component links. We observe that for all of the target capacities the cost of a single link is higher than if we were to aggregate multiple links. We also see that the baseline cost for equipment leads to diminishing benefits for multiplicity as the number of links grows beyond a certain point; for example, overall cost begins to slightly increase beyond five parallel links in the 1 Gbps case. We expect such results exist with any capacity given enough component links. Overall, it appears that we are able to reduce capex with roughly four component links for all of the target capacities, with the largest cost savings at higher capacities.

\begin{figure*}[t]
	\centering
	\begin{subfigure}[t]{0.32\textwidth}
		\includegraphics[width=\textwidth]{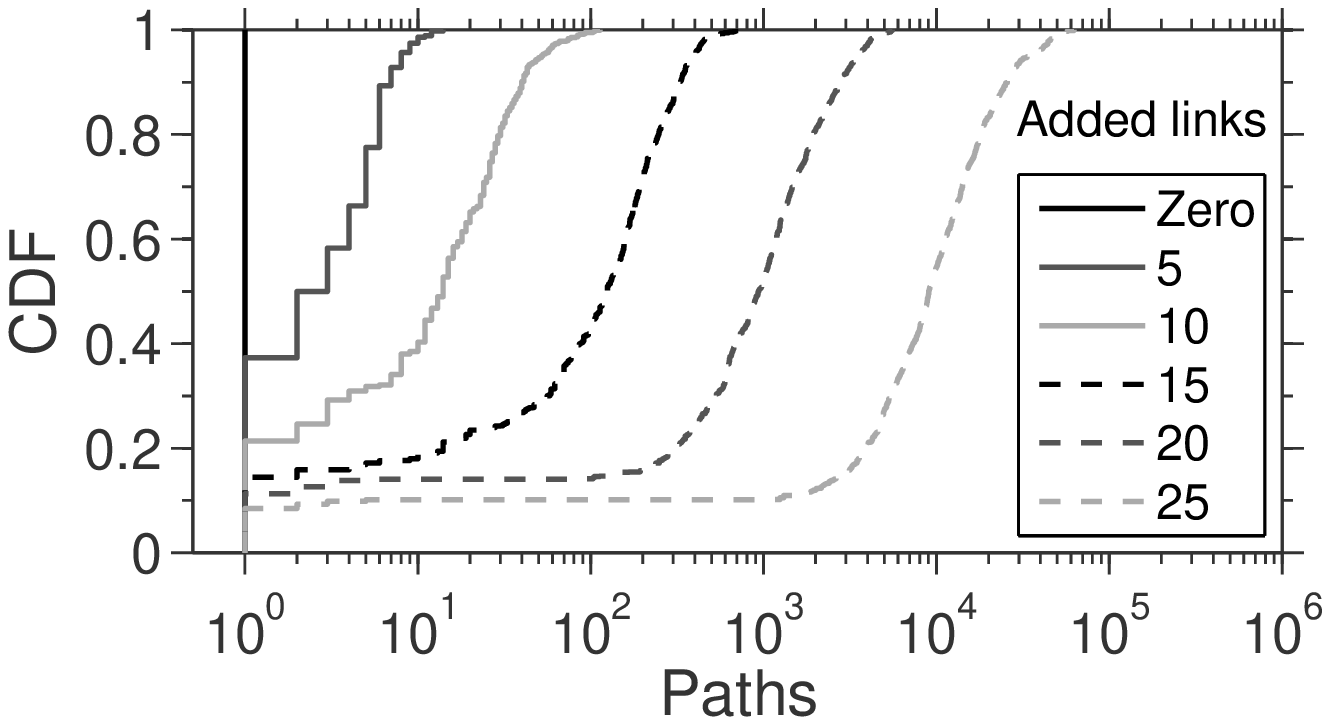}
		\caption{Cross.}
		\label{fig:crosspaths}
	\end{subfigure}
	\hfill
	\begin{subfigure}[t]{0.32\textwidth}
		\includegraphics[width=\textwidth]{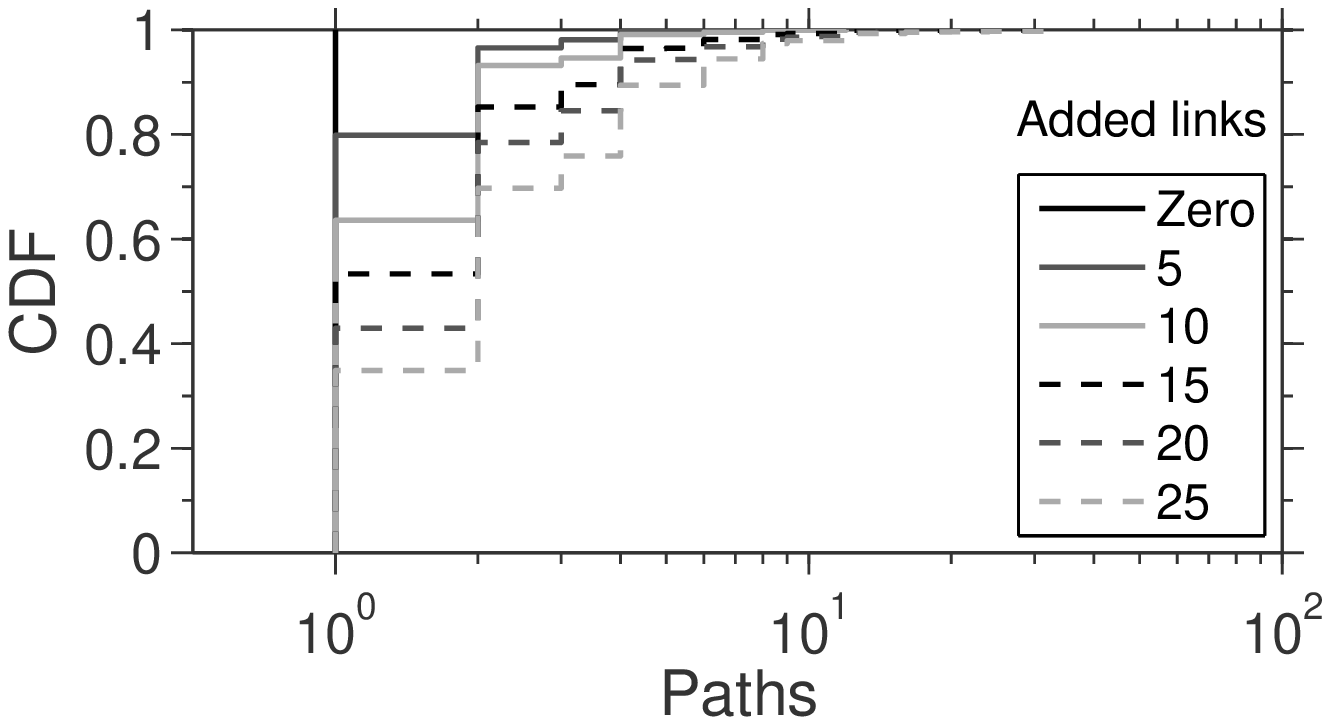}
		\caption{Parallel.}
		\label{fig:parallelpaths}
	\end{subfigure}
	\hfill
	\begin{subfigure}[t]{0.32\textwidth}
		\includegraphics[width=\textwidth]{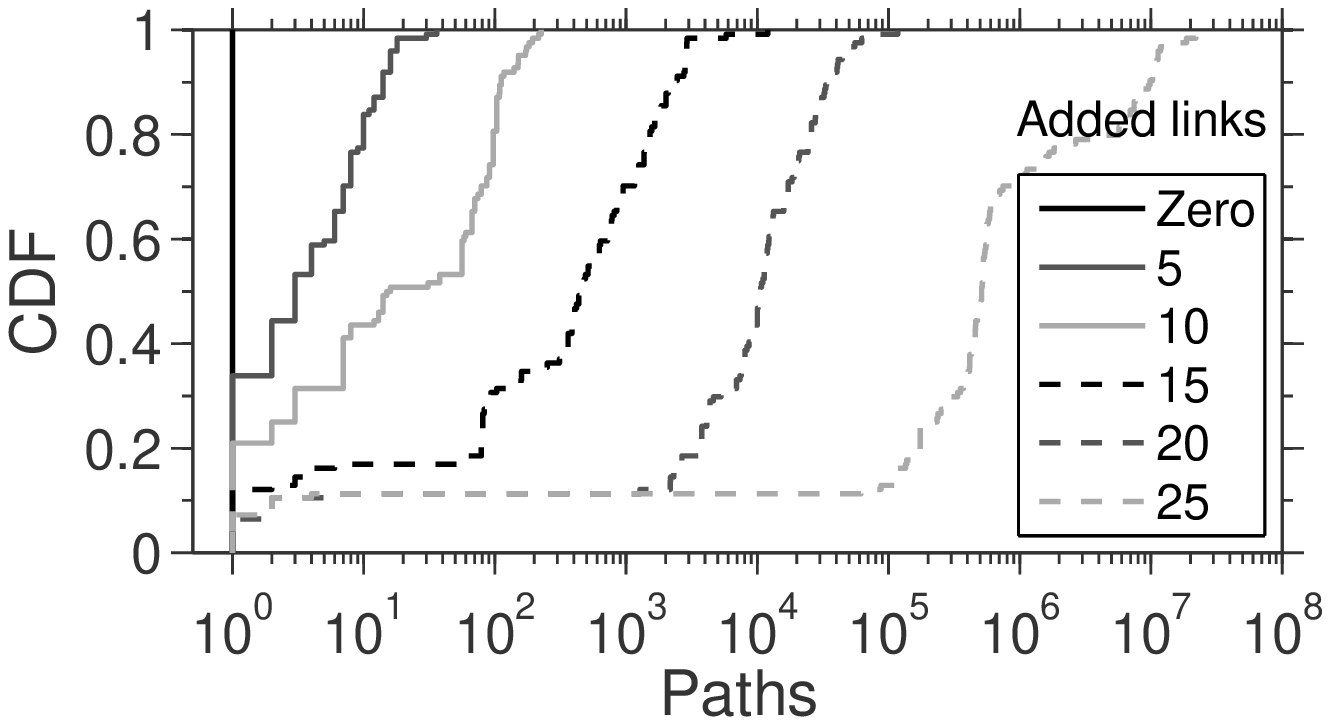}
		\caption{Both.}
		\label{fig:bothpaths}
	\end{subfigure}
	\hspace{\fill}
  \vskip 1em
	\caption{Paths from edge nodes to gateway as parallel, cross, or an equal number of both parallel and cross links are added.}
  \vskip -1em
\end{figure*}

\subsection{Path multiplicity simulation}
\label{sec:simulation}

We explore the benefit of WISPR using a topology from the operational WISP network shown in Figure~\ref{fig:topo}. We omit wired links, and simplify the topology to contain only backhaul nodes and links.  This simplified graph contains 64 nodes (one of which is the upstream gateway) connected via 63 wireless links. There are 41 ``edge'' nodes, which serve as base stations for equipment installed at user sites.

The current network topology, like many WISP networks, is a tree. Thus there is only a single path from each edge node to the Internet gateway. To understand the potential impact WISPR, we conduct three types of analysis.  First, we analyze the multiplicity that is added to this real topology, and as a result, the explosion of the number of available paths.  Second, we consider how we might the minimize cost of the network.  Third, we consider how we might maximize the capacity given a fixed budget.

To compute the growth in path multiplicity we consider three types of new links.  First, we add ``cross'' links between nodes, which connect nodes that are not adjacent in the original topology.  After adding cross links to random node pairs we calculate the number of paths available between each edge node and the Internet gateway.  As we are selecting from among many paths, we must grapple with the scale of multipathing that is made available by the additional links we add.  The resulting CDFs are shown in Figure~\ref{fig:crosspaths}.  Second, we add ``parallel'' links between nodes, which are between a randomly selected node and its upstream neighbor in the original topology.  We again calculate the number of paths between edge nodes and the Internet gateway (Figure~\ref{fig:parallelpaths}).  We observe that, compared with cross links, parallel links tend to result in a lower number of paths to the gateway.  Third, we add the same number of both cross and parallel links to the original topology.  Figure~\ref{fig:bothpaths} shows that the number of paths rapidly grows as links are added.

\subsection{Network cost and capacity simulation}

 As shown in Table~\ref{tab:cost_perf}, the original network topology has a min cut of 2.5 Gbps.  Using price data we discussed in Section~\ref{sec:price}, we estimate the cost of the this subset of the network based solely on the hardware used to be \$27,565.
Using our regression we can estimate the cost to build the same topology with the same capacity using multiple links, while ensuring we do not exceed available spectrum (e.g., 24 20Mhz wireless channels or 11 40MHz channels) at any site.  With the formula we are able to save \$4,510.44 (16.4\%) while using 33 more links than the current network.  Alternatively, since the formula is an estimate backed by real-world gear that exhibits better price/performance ratios, we can use data on that specific equipment.  We are able to re-create the topology with the same capacity by adding 82 additional links while saving \$8,432 (saving 30.6\%), as shown in Table~\ref{tab:cost_perf}.

\begin{table}[t]
	\centering
	\begin{tabular}{|l|c|c|r|}
		\hline
		\textbf{Description}	& \textbf{Capacity}	& \multicolumn{1}{c|}{\textbf{Cost}}	\\ \hline\hline
		Base topology	& 2,500 Mbps	& \$27,565	\\ \hline 
		Fixed capacity, min cost	& 2,500 Mbps	& \$19,133 \\ \hline 
		Fixed cost, max capacity	& 12,700 Mbps	& \$27,353 \\ \hline 
	\end{tabular}
  \vskip 1em
	\caption{Achievable cost savings or capacity gain.}
  \vskip -1em
	\label{tab:cost_perf}
\end{table}
We estimate the potential increase in network capacity using WISPR.  We recreate the original topology; but we replace the equipment throughout with the equipment that offers the best price / performance ratio among commodity hardware.  Links that were previously 100 Mbps or less are replaced with 100 Mbps links.  ``Core'' links connecting sites that are near the gateway or serve a large number of other sites are replaced with 1.4 Gbps links.  All other links in the network are replaced with 400 Mbps links.  Beginning with this baseline tree topology, we then calculate the cost difference between our newly redesigned network and the original topology.  We find that the difference affords us the ability to add 23 additional 400 Mbps links.  We select the 23 new links by iteratively computing the min cut of the network and adding a link that bridges the cut, subject to line of sight and spectrum constraints.  Through this we calculate that, as shown in Table~\ref{tab:cost_perf}, that we can achieve an over 5-fold increase in network capacity for the same cost as the original topology.  However this increase is theoretical; a protocol is needed to actually leverage this new path multiplicity and additional capacity. This has lead us to design WISPR.

\section{System design}
\label{sec:design}
Thus far we have made the demonstrated the need to strengthen WISP networks due to their inherent fragilities (e.g. tree topologies, wireless instability). These factors lead us to make the case for WISPR; a system that allows for the introduction of redundant links, and arbitrary topologies, in WISP networks. In this section, we describe the major components of WISPR, detailing design trade-offs and our approaches for flexibly offering improved performance, improved reliability, or some combination of the two.

WISPR must enable the selection of multiple paths through the network as close to the edge as possible.  Given our requirement to avoid modifying devices or software at customer sites, we treat base stations, which serve as aggregation points for a small number of customer-facing links (dozens, at most), as the network edge.  WISPR must enable multiplicity from this edge to the upstream Internet gateway.  To achieve this we add WISPR nodes---low cost commodity switches with the ability to perform custom forwarding operations---at each major network site.\footnote{In our initial experiments, commodity switches that cost less than \$40 can easily serve as WISPR nodes in that they can forward at line rate in software via four gigabit ports.}  We do not require the modification of existing commodity wireless hardware, and thus in this sense WISPR is not specific to wireless networks (though it proves particularly useful in this context).

WISPR nodes are then responsible for leveraging multiplicity throughout the network.  As WISP networks typically employ legacy routing protocols (such as Spanning Tree), since this is all that many commodity devices support, we cannot rely upon existing support for flexible routing.  Instead, WISPR nodes must become responsible for both control and data plane functionality.  Thus, the WISPR architecture consists of commodity wireless links, which are solely responsible for MAC and physical layer functionality, and the WISPR nodes that serve as intermediaries.

\subsection{Requirements}
After carefully considering the needs of the deployment setting of WISP networks, we converged on the following requirements for the WISPR protocol:\\[0.5ex]
\noindent \textbf{Performance-reliability tradeoff.} The protocol must enable the flexible choice, as in RAID, of leveraging path multiplicity to increase performance, increase reliability, or some combination of the two.\\[0.5ex]
\noindent \textbf{Latency-free.} The protocol must not introduce additional latency to network traffic beyond the delays inherent in the underlying network paths (i.e., any standard propagation, transmission, queuing, and access delays).\\[0.5ex]
\noindent \textbf{Incrementally-deployable.} The protocol must be deployable in WISP networks without requiring any changes at customer sites (e.g., to end device software or hardware, or to gear mounted at customer sites) and also without requiring major changes to commodity devices.\\[0.5ex]
\noindent \textbf{Low cost.} The protocol must not incur significant costs, either computational or financial, to achieve this additional functionality.  Thus we must use only commodity hardware, which are virtually always the lowest cost options.\\[0.5ex]
\noindent \textbf{Backbone-oriented.} The protocol must be designed to work on the network backbone as opposed to being end-to-end, so that the benefits can be directly quantified by the operator rather than depending upon end-to-end behavior.\\[0.5ex]
\noindent \textbf{Asymmetric.} The protocol must be designed to support asymmetric configuration.  That is, the protocol should be able to provide greater performance in one direction (e.g., to improve end-user download performance) and/or greater reliability in another direction (e.g., to improve the reliability of the ACK path).\\[0.5ex]
\noindent \textbf{Adaptive.} The protocol must adapt to current conditions, which can change frequently in wireless environments.  These conditions may be due to user traffic, weather conditions, interference, or other factors.  Given our reliance on commodity hardware deployed in potentially harsh environments, we cannot expect to tease apart the causes of link degradation, but the protocol must still be able to adapt.\\[1ex]
As mentioned earlier, there are many prior multipath solutions, and we originally aimed to leverage one of them. However, after defining the requirements above we found that no prior protocol would meet our needs.  Thus below we sketch the design of a protocol that enables the flexible, asymmetric choice of performance or reliability, does not introduce extra per-packet latency, is incrementally deployable at low cost, and is designed to transparently enhance the backbone of the network.

To provide intuition on our choice to design our own approach, we briefly examine a few potential alternatives. No prior multipath protocols, to our knowledge, support a tradeoff of performance and reliability. One popular choice for pure multipathing is MPTCP~\cite{multipathtcp11}. However MPTCP is designed to replace TCP on a per-flow basis, and thus would require support on user devices (and visibility into path multiplicity at that point). Alternatively, we could create many MPTCP tunnels that begin at the network edge (i.e., base station sites that terminate many user-facing links), but this aggregation is likely to lead to head-of-line blocking and complex, undesirable interactions of MPTCP with the TCP congestion control of tunneled flows (and with UDP-based protocols). In addition, as we describe in Section~\ref{sec:measure}, the overwhelming path diversity that our topology redesign would present would make it very difficult for MPTCP to gain enough knowledge about any individual path choice to effectively converge. Lastly, as we describe in Section~\ref{sec:eval}, MPTCP performance suffers when path loss and latency diverge. Some multipathing systems also attempt to leverage network coding schemes to improve capacity and cope with unreliable links, though such coding typically induces significant delay and is best done on a far coarser granularity than per-packet.

In many networks---such as datacenter networks---ECMP (and variants such as WCMP) provide the ability to leverage hardware support to distribute traffic across parallel paths.  Unfortunately this approach cannot be used in our context for two reasons: the paths available are not equal cost (and thus forwarding in parallel along non-equal cost paths could cause looping) and because we require the ability to flexibly select a performance-reliability tradeoff.  This latter issue is shared by traffic engineering systems for backbone networks, though such systems can select non-equal cost paths to spread traffic and avoid network hotspots.

\subsection{Constraints}
As we intend for WISPR to be a generalizable solution for WISP networks, we must consider the limitations of common WISP network equipment:\\[0.5ex]
\noindent \textbf{Heterogeneous hardware support.} WISP networks are often built out incrementally as new customer service areas are identified. Likewise, WISPs commonly utilize components from multiple vendors and with varying capacities. WISPR must operate across a wide range of equipment.\\[0.5ex]
\noindent \textbf{Hardware limitations.} Our design requires the addition of WISPR nodes to provide multipath-enabled ingress and egress for user traffic across the WISP network. We engineer WISPR to operate on commodity switch / router devices. Such low-cost hardware typically has rather low processor and memory specifications. As such, WISPR must be frugal with processing resources.\\[0.5ex]

\subsection{WISPR Levels}

For concreteness we present a prototype design to achieve a range of performance and reliability tradeoffs. Inspired by RAID, WISPR provides a similar model of different `levels' of reliability through the use of mirroring, striping, and parity. WISPR levels and their specific parameters can be selected on a per node-pair basis, with upstream and downstream directions chosen independently. This enables, for example, the selection of high performance (e.g., WISPR 0) in one direction and high reliability (e.g., WISPR 1) in another. Given our observations of the WISP in Section~\ref{sec:measure}, we implement striping and mirroring in our prototype as we did not witness high prevalence of packet corruption in the SNMP traces, which would motivate parity-related WISPR levels. Such extension is left for future implementation.

\subsubsection{WISPR 0}
Packets are striped across available paths between source and destination WISPR nodes. This mode offers maximally increased performance compared to the use of a single path. WISPR 0 offers no redundancy or parity. Link failures, manifesting as packet loss, are ignored in WISPR 0 and are addressed by higher layer protocols, though they affect path weights.  There are a number of key details to enabling striping across paths in the network; here we present a preliminary exploration of this design.

As we discusse in Section~\ref{sec:simulation}, when we evaluate target network topologies and redesign them with the WISPR approach we make available a huge number of potential paths from the edge to the upstream gateway.  This presents two immediate protocol constraints.  First, weighting traffic across potential paths (given the diversity of path capacities) requires aggregating knowledge about paths on the basis of their underlying links rather than on an end-to-end basis, as no path will be used with sufficient frequency to collect useful capacity statistics.  Second, WISPR must leverage many paths that overlap both in nodes and links, so the protocol must account for this overlap in path selection.

As a result of these constraints, we must disaggregate path characteristics as packets traverse different paths and arrive at the destination WISPR node and distribute the underlying link (and thus path) information to the WISPR nodes so they can make appropriate selections of paths. Our preliminary design of WISPR 0 involves the logically-centralized collection of topology information using an SDN controller hosted at the upstream gateway and the distribution of path information in the form of OpenFlow rules that are weighted according to the capacity of paths.  We plan to extend Open vSwitch weighted multipath support we built previously to achieve appropriate data-plane support at WISPR nodes.  However since we do not wish to impose a large number of rule changes for every change in weights, we instead translate the weights on paths into weights on outbound ports on a switch-by-switch basis.

However the constraints above present an additional challenge: we wish to ensure that paths are selected that are maximally disjoint~\cite{lee2001split}, so that we can provide good performance guarantees.  Selecting disjoint paths at WISPR nodes on the data plane would be prohibitively expensive.  Instead, our approach is to compute disjoint paths at the controller and group them such that they are used in parallel for a short period of time.  Due to the large number of paths available, we have the ability to select the multiplicity $k$ of WISPR in all modes---for example, we might choose a cluster of 5 paths across which we stripe at one timestep.  This keeps flows pinned to paths and simultaneously ensure that new flows are placed on disjoint paths.  Due to the randomness involved both in flow arrival and in weighted selection of paths, we cannot compute the huge number of possible disjoint sets in advance regardless of order of flow-path placement.  Configured parallelism enables us to provide predictable performance and path diversity.

\subsubsection{WISPR 1}
In WISPR 1, packets are mirrored across multiple paths. Since the number of paths available now is dramatically greater, we opt for configurable multiplicity $k$, as with WISPR 0, that indicates the number of parallel paths along which each packet should be mirrored.

RAID 1E provides redundancy by permanently storing the mirrored data on multiple disks. On the other hand, WISPR employs mirroring to increase the likelihood that a packet traverses the network between the sender / receiver WISPR node pair. We do \emph{not} want to forward multiple copies of a mirrored packet upstream. As packets arrive at the receiver WISPR node, they are deduplicated.

\subsubsection{WISPR 4}
WISPR 4 employs dedicated parity paths in addition to striping packets across multiple links.  This is especially useful in our context where links are frequently flaky and/or encounter losses due to wireless issues (including but not limited to weather, interference, and access congestion).
WISPR aims to provide parity across paths that are not equivalent and parallel. As a result, we must resolve the challenge of both group formation and parity generation.

\noindent 

\textbf{Parity.} We enable parity by placing packets in a group, at the specified level of multiplicity $k-1$ (to allow for a single parity). We chose an arbitrary number $X$ and say that every $X$th packet will be parity packet. The group of $X$ packets is striped as in WISPR 0. Each time a packet is emitted out a given port, the packet counter is incremented. The parity packet represents the XOR of the previous $X-1$ packets, and is generated after $X-1$ packets have been sent. Both the ingress and egress nodes will have have the value of $X$ to identify the parity packets.

\subsubsection{WISPR 5}
WISPR 5 is equivalent to WISPR 4 except the parity packets rotate across paths.

\section{Implementation}
We implement a prototype of WISPR to evaluate its effectiveness. We target OpenWRT\footnote{\url{https://openwrt.org/}} as the OS platform for our system due to it's broad support for commodity networking hardware. The controller is based on OpenFlow and we utilize Open vSwitch\footnote{\url{http://openvswitch.org/}} (OVS) as the basis for our forwarding policies as it is well supported by OpenWRT. WISPR includes three high-level components: a software-defined networking-based controller, the control plane, and the data plane.  We detail the design choices and trade-offs for each in the following sections.

\subsection{Controller}

We use a RYU controller to get the entire network topology, to find available multiple paths between the source and upstream gateway, and to install rules on OVS switches to route packet on different paths.

\subsection{Data Plane}

 \subsubsection{WISPR daemon}

 A WISPR-aware daemon runs on all the edge nodes that participate in our protocol. The daemon processes ingress and egress traffic when the node is one of the endpoints in a WISPR session. Ingress frames are encapsulated with a VLAN header and the vlan-id field is used to store WISPR session state. At the egress node the session state is read from the header and used for processing. We have chosen to initially implement the daemon as a userspace process running in OpenWRT for flexibility in development and experimentation. We anticipate moving the system to a kernel module in the future in order to avoid costly overheads associated with userspace processes. The daemon is written in C for portability.

\subsubsection{Node-pair WISPR sessions}

 The WISPR session state, the mode, and options associated with the session such as path identifier can be associated with traffic between any two WISPR nodes in the network. Importantly, this includes WISPR pairs at the edge of the network for cases of intra-network communication as well as traffic destined to and from a network internet gateway. For each session, we again rely on VLAN encapsulation between the WISPR node ingress-egress pair. The reason we use VLAN encapsulation is twofold: 1) it allows us to efficiently move frames toward their intended destination (i.e. quickly passing over interim WISPR nodes that are neither the ingress or egress point), and 2) we leverage the 12-bit vlan-id space present in the VLAN header to include session state information to minimize our overhead footprint. The 12-bit vlan-id space is not enough to hold session information, so we use 2 VLAN headers to increase the space to 24 bits. In the future when GRE rule matching functionality is available the VLAN headers can be replaced with one GRE header and 32-bit 'key' field can be used to hold session information.

\subsubsection{Session state}

 WISPR nodes need information identifying WISPR traffic paths and the destination. We want to use space in real headers - we use VLAN encapsulation and 12 bit vlan-id field. We use 2 vlan headers to increase the bit space to 24. 4 bits for WISPR mode, 16 bits for packet id, and 4 bits for path identifier. 

\begin{figure}[t!]
	\centering
  	\includegraphics[trim=0 0 0 0, clip=true,width=\columnwidth]{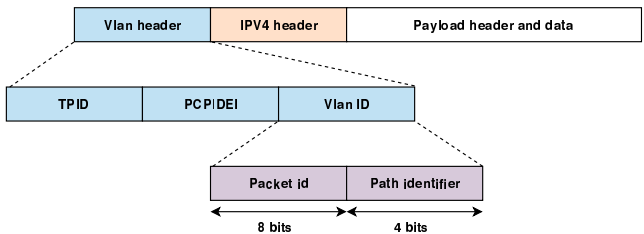}
  	\caption{WISPR state is stored within the VLAN's vlan\_id field.}
  	\label{fig:wisprheader2}
\end{figure}

\textbf{WISPR path index}. The 4-bit path index is used to identify a path for a packet by the WISPR node. The egress node increases the index value after sending each packet along a different path. The index value resets to 0 upon reaching the maximum number of paths.

\textbf{WISPR packet id}. The 16-bit bitmask field is used to identify packets. At the ingress node, the packet id value is incremented for every new packet. 

For parity packets, we chose an arbitrary number $X$ and every $X$th packet will be a parity packet. Both ingress and egress node will have the value of $X$.

\subsubsection{Deduplication}
WISPR mode 1 mirrors traffic across multiple paths between WISPR nodes. Accordingly, the egress node must deduplicate mirrored packets to avoid wasting bandwidth beyond the boundaries of the WISPR session. There are multiple locations where packet deduplication could be accomplished: 1) the controller, 2) Open vSwitch, and 3) the WISPR daemon. We investigated all three options and found the daemon to be the most flexible choice. Using the centralized controller to deduplicate packets would be highly inefficient as all traffic would need to pass through the controller. Open vSwitch, in its current state, does not have packet deduplication functionality. If we added deduplication to OVS, switching performance for all traffic through the node would suffer due to the processing overhead. By using the daemon, we introduce a performance hit only on the packets that must be processed at the node, all other packets can pass through the node without delay. 

For duplication of the packet, egress node increments the path index field and keep the packet number the same. The WISPR daemon creates a hash table for ingress WISPR sessions and observes as packets arrive. The first copy of any mirrored packet that arrives at the egress node is deencapsulated (i.e. stripped of the VLAN header) and the payload is forwarded on for regular IP routing toward the destination. The packet id is placed in the hash table for future reference. Subsequent copies of the packet are checked against the hash table, if the packet has been seen before it is simply dropped by the daemon. The hash table is a circular array, thus mirroring sessions are limited in terms of the number of packets in-flight between nodes to $ 2^{16} = 65,536 $.

 Packets may arrive out of order due to multiple paths. To tackle this problem we use a window (similar to TCP's sliding window). We use four equal sized windows and a index pointer rather than using sliding window because we do not get ACKs/NACK as in TCP. When a packet is received, if the existing index pointer is less than the packet id, the index pointer is set to the new packet id. Otherwise, since the packet id is less than the index pointer, the packet id is checked to see whether it falls within the current or previous windows. If the packet is outside of these two windows it will be considered a duplicate and discarded. Wraparound occurs when index pointer reaches the maximum range of window. WISPR can support maximum bit rate of 32 Mbps for wraparound to occur in 1 sec given that size of all packets is 64 bytes (minimum packet size) [64 (minimum packet size) * 65,536 (maximum window size) = 32 Mbits].

\subsection{Discussion}
While link-layer retransmission can mask losses they introduce significant latency, and on backhaul links can cause head-of-line blocking.  Many WISP network operators decrease MAC-layer retransmissions or turn them off entirely.  Nedevschi found that a retransmission limit of one was ideal in such networks~\cite{Nedevschi:EECS-2009-27}.  We are agnostic to this choice---retransmissions on a per-link basis will make individual path segments appear to be less lossy, but introduce higher delay.
\section{Evaluation}
\label{sec:eval}

\begin{figure}[t!]
	 \centering
  	 \includegraphics[trim=0 0 0 0, clip=true,width=1\columnwidth]{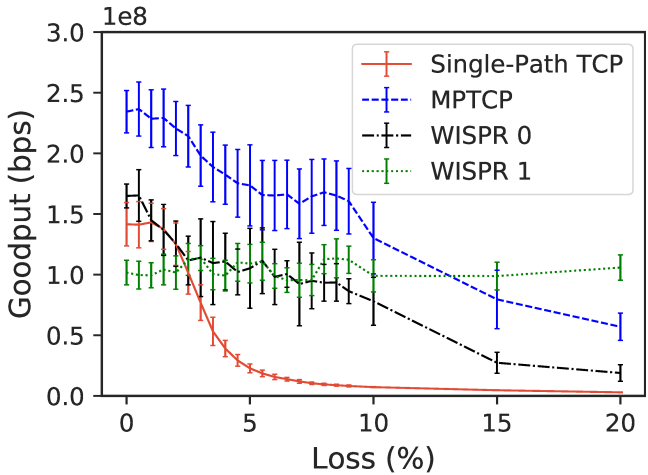}
  	 \caption{WISPR goodput.}
  	 \label{fig:goodputwisprnew}
 \end{figure}

\noindent
\subsection{TCP Goodput}
We evaluate TCP goodput in a lab environment. We construct a network with two paths between hosts, two WISPR nodes (one adjacent to each host), and an inline linux machine that allows us to introduce packet loss using \texttt{tc} on one of the paths. To understand the benefits of WISPR and compare with MPTCP we find the goodput of a regular, single-path TCP flow running over the lossy path using \texttt{iperf} (red curve in Figure~\ref{fig:goodputwisprnew}). 
We observe that
MPTCP significantly outperforms regular TCP and provides high goodput until higher loss scenarios (i.e. 15\% and 20\% loss). 
Furthermore, as loss increases, the goodput achieved by MPTCP is also better than WISPR 0 and WISPR 1, until high loss environments.
When the loss rate is low, WISPR 1 underperforms regular TCP and MPTCP, which we attribute to the processing overhead of our daemon running in userspace. However, as the loss rate increases, WISPR 1 goodput remains relatively stable whereas regular TCP and MPTCP decay quickly. In high-loss networks, a common scenario in WISP networks, WISPR 1 manages to provide stable goodput.

\noindent
\subsection{TCP Delay}
To understand how WISPR would perform on a more realistic network topology, we use the Mendocino measurements in Section~\ref{sec:measure} to synthesize a topology with multiple paths and hops. This dataset contains the geographical coordinates of the sites, but not the links in the network. To simulate the existence of multiple paths, we introduce links to connect the nodes as follows: First, we use Zyxt~\cite{potsch2018zyxt} (a network planning tool) to cluster nodes in a radius of 500m into 1 node to simulate aggregation of customer sites through an omnidirectional access point. Then, we use breadth-first search starting from the upstream gateway node (the center-most node) to construct a hierarchical ordering. Starting from leaf nodes in this hierarchy, we connect the nodes in such a manner that each node at a depth $d$ is randomly connected to up to 10 nodes at depth $d-1$ in hierarchy. We construct 5 shortest node disjoint paths between hosts using python NetworkX library. Using this topology, we measure average delay of data transfer between MPTCP, single path TCP and WISPR. In this set of experiments we focus on WISPR 1 (striping and mirroring).

\begin{figure}[t!]
	\centering
  	\includegraphics[trim=0 0 0 0, clip=true,width=1\columnwidth]{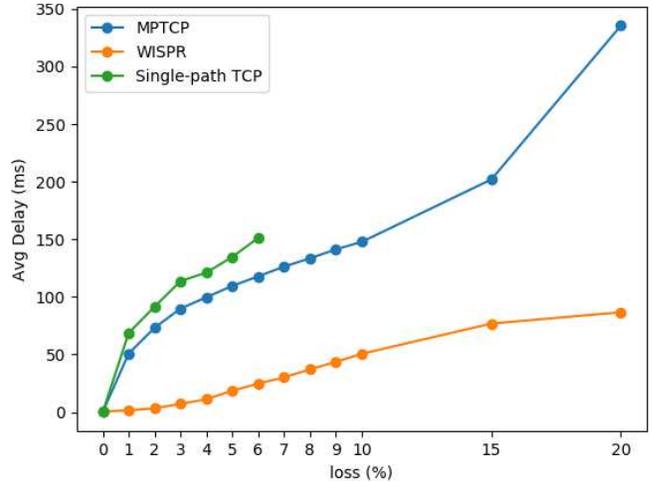}
  	\caption{Loss - singlePath TCP Vs WISPR Vs MPTCP 50,000 packets.}
  	\label{fig:lossmptcpVsWISPR50000}
\end{figure}

We tune the percent of packet loss using OVS switch API for \emph{all} 5 paths. To understand the benefits of WISPR, we compare the delays of single-path TCP, MPTCP, and WISPR. We use a C program to generate packets of 96 bytes with a sequence number and timestamp in the packet payload. At the receiver host we get timestamp from incoming packet and the current time when packet was received, then measure the time difference (delay time) between the two timestamps. We introduce a delay of 1,000 microseconds between sending 2 packets so that packets do not get dropped due to congestion. We keep the bandwidth constant for all paths (10Mbps). Figure~\ref{fig:lossmptcpVsWISPR50000} shows the results after sending 50,000 packets ($\approx$5MB);
WISPR performs better than MPTCP in terms of average delay as we increase the packet loss rate across the network. Single path TCP performs poorly due to retransmissions and until eventually the connection is timed out at higher loss rates.

\noindent
\subsection{UDP Loss and Delay}
In this experiment, we compare the results of running WISPR for UDP traffic over a high delay and lossy network as is common in WISP networks. We simulate a single path through our Mendocino topology and introduce a delay of 500 ms (to simulate weather or congestion-related delays) and packet loss of 5\% between 2 hosts for both original tree topology and WISPR using OVS switch API,
then measure the packet loss and average delay. We run this experiment for 10 iterations and send 500 packets per iteration. We find that UDP produces an average of 26.6 packets lost compared to WISPR, which loses 0 packets. Furthermore, the average UDP delay is 530 ms compared to 18 ms for WISPR.

\section{Related Work}

In years past there has been important work in a number of areas of wireless networking that are orthogonal but complementary to WISPR. Specifically, in this paper we are not concerned with wireless mesh networks, ad-hoc networks, or sensor networks, all of which have been prominent in the literature over the span of two decades~\cite{aguayo2004link,bicket2005architecture,biswas2005exor, johnson96dsr, perkins99aodv, levis2004trickle, werner2006deploying, ganesan2001highly, estrin99challenges,intanagonwiwat00diffusion}.  Such networks, while interesting in their own ways, are technologically, topologically, administratively, and economically different from WISP networks.  Similarly, we are not focused on enterprise or campus wireless networks, which are also very different in that they typically provide connectivity via omnidirectional APs but are directly connected to wired upstream bandwidth~\cite{meraki,aruba}.  In addition, there is significant work being done on extending the paradigm of wireless at lower layers, leveraging the power of software to both improve performance~\cite{katti06xors,katti2007embracing,gollakota2008zigzag} and unlock surprising new functionality~\cite{adib2013see,liu2013ambient}.  Finally, while cellular networks are complementary to providing the type of network access we are interested in (and we have done work on them in parallel to this effort), the technical and economic challenges of such networks are very different, as are the regulatory constraints~\cite{heimerl2013local,Schmitt:2016:NCR:2926676.2926691}.

There are major R\&D efforts underway at Facebook and Google, among others, to use drones~\cite{fbdrone} or balloons~\cite{loon} to provide rural connectivity, though these efforts are best at providing certain types of connectivity (e.g., middle mile) in certain types of regions.  Indeed, despite its ongoing drone R\&D, Facebook has partnered with a number of WISPs to expand connectivity, and Google has funded WISP efforts as well.  We similarly do not see these as mutually exclusive approaches to providing rural connectivity, and in fact there are opportunities to combine WISP networks with cell networks and aerial connectivity to build hybrid networks that leverage the advantages of each.

In the context of datacenter networks, there have been many innovative designs for leveraging heterogeneity and multipathing to improve performance and reliability, including the use of wireless links~\cite{liu2013f10,liu2015subways,zhou2014wcmp,halperin2011augmenting,zhou2012mirror,hamedazimi2015firefly,hedera10}.
Often in such contexts techniques such as link aggregation are used; several vendors have extended the concept of link aggregation protocols such as LACP~\cite{LACPstandard} to multi-chassis LACP.  Such protocols are useful in narrow settings with device homogeneity and rigid network structure.

The three classes of prior work that are most relevant, but ultimately different from WISPR, are multipath protocols~\cite{multipathtcp11,De2003481,Godfrey:2009:PR:1594977.1592583, Ganichev:2010:YYM:1880153.1880156, miro,deflections,ganesan2001highly}, traffic engineering designs for carrier networks~\cite{elwalid2001mate, kandula2005texcp, wang2006cope}, and multipath network coding schemes~\cite{netcoding,katti06xors}.  
These protocols do not meet our needs because none of these provide a combination of an in-network view of lossy links (which are common in WISPs, unlike carrier networks), dynamic path selection, low latency, and the ability to flexibly choose a tradeoff between reliability and performance.

Finally, in the context of WISP networks specifically, a recent work by Potsch et al.~\cite{potsch2018zyxt} focus on how to design and plan for the large scale deployment of rural wireless networks. WISPR synergizes with that work by providing a protocol that can leverage multiple paths to enable more robust network designs.

\section{Conclusion}
The most viable and cost-effective approach for providing last-mile connectivity has proved to be Wireless ISPs (WISPs), which rely on point-to-point wireless backhaul infrastructure to provide connectivity using cheap commodity wireless hardware. However, unlike conventional infrastructures, individual WISP network links are known to have poor reliability due to a wide variety of environmental factors. Furthermore, from a cost analysis of existing wireless hardware trends, we show that WISP commodity wireless hardware has an inherent cost-performance tradeoff that can be leveraged to move toward a greater number of inexpensive links thereby lowering costs and increasing redundancy. Since no good existing solutions exist to take advantage of this new path diversity, we introduced Wireless ISPs with Redundancy (WISPR). We demonstrate that WISPR is able to provide increased performance, reliability, or a combination of the two in WISP networks where unpredictable links are the norm. Increasing performance and reliability while lowering cost makes WISP networks more economically viable and thus extends their potential reach into rural areas.

\bibliographystyle{ACM-Reference-Format}
\bibliography{bib}


\end{document}